\documentclass[prl,twocolumn,superscriptaddress]{revtex4}
\usepackage[dvips]{graphicx}
\usepackage{latexsym}
\usepackage{bm}
\newcommand{\fig}[2]{\includegraphics[width=#1]{#2}}

\begin{document}

\renewcommand{\ni}{{\noindent}}
\newcommand{\dprime}{{\prime\prime}}
\newcommand{\be}{\begin{equation}}
\newcommand{\ee}{\end{equation}}
\newcommand{\bea}{\begin{eqnarray}} 
\newcommand{\eea}{\end{eqnarray}}
\newcommand{\nn}{\nonumber} 
\newcommand{\bk}{{\bf k}}
\newcommand{\bQ}{{\bf Q}}
\newcommand{\bN}{{\bf \nabla}}
\newcommand{\bA}{{\bf A}}
\newcommand{\bE}{{\bf E}}
\newcommand{\bj}{{\bf j}}
\newcommand{\bJ}{{\bf J}}
\newcommand{\bs}{{\bf v}_s}
\newcommand{\bn}{{\bf v}_n}
\newcommand{\bv}{{\bf v}} 
\newcommand{\la}{\langle}
\newcommand{\ra}{\rangle} 
\newcommand{\dg}{\dagger}
\newcommand{\br}{{\bf{r}}} 
\newcommand{\brp}{{\bf{r}^\prime}} 
\newcommand{\bq}{{\bf{q}}}
\newcommand{\hx}{\hat{\bf x}} 
\newcommand{\hy}{\hat{\bf y}}
\newcommand{\bS}{{\bf S}} 
\newcommand{\cU}{{\cal U}}
\newcommand{\cD}{{\cal D}} 
\newcommand{\bR}{{\bf R}}
\newcommand{\pll}{\parallel} 
\newcommand{\sumr}{\sum_{\vr}} 
\newcommand{\cP}{{\cal P}} 
\newcommand{\cQ}{{\cal Q}} 
\newcommand{\cS}{{\cal S}}
\newcommand{\upa}{\uparrow} 
\newcommand{\dna}{\downarrow}
\def\gtrsim{\stackrel{_>}{_\sim}}

\title{Nodal Quasiparticle Dispersion in Strongly Correlated
d-wave Superconductors}
\author{Mohit Randeria}
\affiliation{Department of Physics, University of Illinois at
Urbana-Champaign, IL 61801}
\affiliation{Department of Theoretical Physics, Tata Institute of
Fundamental Research, Mumbai 400 005, India}
\author{Arun Paramekanti}
\affiliation{Department of Physics and
Kavli Institute for Theoretical Physics, University 
of California, Santa Barbara, CA 93106--4030}
\author{Nandini Trivedi}
\affiliation{Department of Physics, University of Illinois at 
Urbana-Champaign, IL 61801}
\affiliation{Department of Theoretical Physics, Tata Institute of 
Fundamental Research, Mumbai 400 005, India}
\begin{abstract}
We analyze the effects of a $\bk$-dependent self-energy on the
photoemission momentum distribution curve (MDC) dispersion and
lineshape. We illustrate this general analysis by a detailed 
examination of nodal quasiparticles in high Tc cuprates.
Using variational results for the nodal quasiparticle $Z$, which varies
rapidly with hole doping $x$, and $v_F^{\rm low}$, which is independent of $x$,
we show that the high energy dispersion $v_F^{\rm high} = v_F^{\rm low}/Z$,
so that it is much larger than the bare band structure dispersion and
also exhibits strong doping dependence in good agreement with
recent photoemission data.
\vspace{0.1cm}
\typeout{polish abstract}
\end{abstract}

\maketitle
Nodal quasiparticles (QP's) are the dominant low-lying excitations
in the superconducting state of the high Tc superconductors, and 
have been the focus of intense experimental and theoretical investigation 
\cite{orenstein00}.
In particular, angle resolved photoemission spectroscopy (ARPES), 
exploiting the simple Lorentzian lineshape of
momentum distribution curves (MDC's) \cite{valla99}, has provided
a great deal of insight into the nodal 
QP weight and dispersion \cite{valla99,kaminski00,bogdanov00,lanzara01,
johnson01,yoshida02,zhou03}. Two important questions are: (i) What
constraints do these experiments place on a model for the superconducting
state? (ii) To what extent are nodal QP properties determined
purely by strong electronic interactions?

The purpose of this Letter is twofold. First, we examine the 
effect of the $\bk$-dependence of the electronic self energy on the
analysis of MDC's for gapless nodal QP's.
Specifically, we determine how this affects the MDC dispersion, 
after clarifying the constraints under which the 
MDC lineshape is Lorentzian. This is important, because the
usual basis for understanding a Lorentzian lineshape is a $\bk$-independent
self-energy \cite{campuzano_review}, which is certainly sufficient but
by no means necessary as shown here.

The second goal is to use the above formalism to gain insight into some
unusual aspects of the nodal QP dispersion in 
the high Tc superconductors.
Some time ago we predicted, on the basis of a 
variational calculation\cite{paramekanti01} for a Hubbard model
that the nodal QP spectral
weight $Z$ vanishes linearly with hole doping $x$ even though
their Fermi velocity $v_F^{\rm low}$ remains doping independent
as $x \to 0$. (We use the notation $v_F^{\rm low}$ here, rather than
the more conventional $v_F$, to clearly 
distinguish it from $v_F^{\rm high}$ defined below). 
These predictions have now been verified in some 
detail by angle-resolved photoemission spectroscopy
(ARPES) experiments \cite{johnson01,yoshida02,zhou03}. 

These same experiments also show an unusual doping dependence
of the {\em high energy dispersion}, at binding energies larger than the 
much-studied kink \cite{johnson01,lanzara01,zhou03,norman,abanov} 
in the nodal MDC dispersion. The ``high energy'' Fermi velocity
$v_F^{\rm high}$, at energies of around 150 to 200 meV, 
is experimentally found \cite{johnson01,lanzara01,zhou03}
to be quite strongly doping 
dependent and {\em increases} markedly with 
underdoping, where as one might have naively expected 
the ``bands'' to become narrower with decreasing $x$. 
This behavior of $v_F^{\rm high}$ is also in marked contrast 
with both the Fermi velocity at $E_F$ which, as stated above, 
is $x$-independent, and the ``bare'' band-structure Fermi velocity 
$v_F^0$ which has a rather weak $x$-dependence. 

In this Letter, we show that this surprising behavior of
$v_F^{\rm high}(x)$ follows directly from the same theory 
\cite{paramekanti01} which predicted $Z(x)$ and
the $x$-independent $v_F^{\rm low}$.
As discussed below, the results for $Z(x)$ and $v_F^{\rm low}$ imply 
strong constraints
on {\em both} the $\omega$-\, {\em and} $\bk$-dependence of 
the real part of the self energy $\Sigma^\prime(\bk,\omega)$.
Specifically, both $\partial\Sigma^\prime/\partial \omega$
and $\partial\Sigma^\prime/\partial k$ must exhibit $1/x$ singularities.
The strong $\bk$-dependence is somewhat unexpected, and 
we show here that it has 
an important effect on  the ``high energy'' dispersion:
\be
v_F^{\rm high} = v_F^0 + \partial\Sigma^\prime/\partial k 
= v_F^{\rm low}/Z
\label{vfhigh}
\ee
leading to a strong doping dependence in $v_F^{\rm high}$,
including an $1/x$ divergence as the hole doping $x \to 0$.
We make detailed quantitative comparison of our results with 
existing ARPES data, and conclude with comments on the kink 
in the MDC dispersion.


\medskip
\ni
{\bf MDC analysis:}
We first develop the formalism for the analysis of photoemission
MDC's, focusing on two
points {\em not} emphasized in the literature: the conditions
under which the MDC lineshape can be Lorentzian despite
$\bk$-dependence of the self energy, and the role of
$\partial\Sigma^{\prime}/\partial k$ in the MDC dispersion.
Recall that the ARPES intensity from a two-dimensional system is given by 
\cite{randeria95}
$I(\bk,\omega) = I_0(\bk)f(\omega)A_{\bk,\omega}$
where $I_0$ contains the dipole matrix element and kinematical
factors, $f(\omega)$ is the Fermi function and $A_{\bk,\omega}$
is the one-particle spectral function. Here $\bk$ is the 2D momentum and
$\omega$ the energy measured from the chemical potential.
The usual energy distribution curve (EDC) method views ARPES data
as a function of energy at fixed $\bk$.
The MDC method, on the other hand, analyzes
$I(\bk,\omega)$ as a function of $\bk$, along a suitable chosen cut in 
$\bk$-space, for fixed $\omega$.   The MDC method is more powerful
than the traditional EDC for studying gapless excitations, because
of the simple Lorentzian MDC lineshape and also the ease
of unambiguously identifying the extrinsic background. 

Here we are specifically interested in the MDC's
for a $d_{x^2 - y^2}$ superconductor for $k$ along the zone diagonal,
so that the ``off-diagonal'' self energy (gap function) vanishes. 
The formalism derived is obviously applicable near the Fermi surface
of any normal metal also.
The spectral function is then given by:
\be
\!~A_{\bk,\omega}\! =\! {1 \over \pi} {{\vert \Sigma^{\prime\prime}(\bk,\omega) \vert}
\over {[\omega - \epsilon_0(\bk) + \mu_0 - \Sigma^\prime(\bk,\omega)]^2
+ [\Sigma^{\prime\prime}(\bk,\omega)]^2}}
\label{akw}
\ee
where $\epsilon_0(\bk)$ and $\mu_0$ are the ``bare'' dispersion
and chemical potential, and $\Sigma = \Sigma^\prime + i \Sigma^{\prime\prime}$
the self-energy.

The ``renormalized'' $k_F$ is defined by the condition
$\epsilon_0(\bk_F) - \mu_0 + \Sigma^\prime(\bk_F,0) = 0$.
In case interactions lead to no change in the $k_F$,
this places a constraint $\Sigma^\prime(\bk_F,0) = 0$
on the real part of the self energy.
We now expand in small $| k - k_F | \ll k_F$ in the vicinity
of the node at $k_F$, keeping $\omega$ arbitrary for the moment.
Linearizing the dispersion, we thus get
\be
\epsilon_0(\bk) - \mu_0 + \Sigma^\prime(\bk_F,0) \simeq v_F^0(k - k_F)
\ee
where $v_F^0$ is called the bare Fermi velocity.

We next assume
$\Sigma^{\prime\prime}(\bk,\omega) \simeq \Sigma^{\prime\prime}(\bk_F,\omega)$
which is equivalent to setting the first correction 
\be
\partial\Sigma^{\prime\prime}(\bk_F,\omega)/\partial k = 0,
\label{constraint}
\ee
for all $\omega$ of interest.
This is a {\em necessary} condition for a Lorentzian lineshape in
$(k - k_F)$, since without it the numerator in eq.~(\ref{akw}) would have
a term linear in $(k - k_F)$ which would make the expression non-Lorentzian.
The Kramers-Kr\"onig transformation of the above condition implies that 
$\partial\Sigma^{\prime}/\partial k$ can only be a constant
independent of $\omega$, a condition we will need to use below.
(In the experiments, an additional constraint for
a Lorentzian MDC is that the matrix element prefactor $I_0$ should be
$\bk$-independent in the range of relevant $\bk$'s.)

With this one assumption, the spectral function $A_{\bk,\omega}$ is
easily shown to be a Lorentzian in $(k-k_F)$ peaked at
\be
k(\omega) = k_F + \frac{\omega -\left[\Sigma^\prime(\bk_F,\omega) - 
\Sigma^\prime(\bk_F,0)\right]}{\left(v_F^0 + \partial\Sigma^\prime/\partial 
k \right)}
\label{dispersion}
\ee
with a width
$\Delta k = {\vert \Sigma^{\prime\prime}(\bk_F,\omega) \vert /
{\left(v_F^0 + \partial\Sigma^\prime/\partial k \right)}}$.
The dispersion of this Lorentzian MDC is then obtained from
\be
\frac{dk}{d\omega} = \frac{1-\partial\Sigma^\prime(k_F,\omega)/
\partial\omega}{v_F^0+
\partial\Sigma^\prime/\partial k} \equiv \frac{\zeta(\omega)}{v_F^0+
\partial \Sigma^\prime/ \partial k},
\label{dispersion}
\ee
where the denominator is $\omega$-independent and evaluated at $\bk\!\!=
\!\!\bk_F$,
but the numerator at arbitrary $\omega$ and $\bk\!\!=\!\!\bk_F$.

In the limit $\omega \to 0$, the MDC dispersion of eq.~(\ref{dispersion}) 
yields the standard result. 
Defining the low-energy renormalized Fermi velocity $v_F^{\rm low}$ via 
$1 / v_F^{\rm low} = dk/d\omega(\omega \to 0)$, we get
\be
v_F^{\rm low} = Z \left[v_F^0 + \partial\Sigma^\prime/\partial k \right].
\label{vflow}
\ee
Here the QP weight $Z$ is defined by
\be
Z \equiv {1 / \zeta(\omega \to 0)}
= {1 / \left[1 -\partial\Sigma^\prime/\partial \omega\right]},
\label{residue}
\ee
with $\partial\Sigma^\prime/\partial \omega$ is evaluated at
$(\bk = \bk_F,\omega = 0)$.

On the other hand, one can quite generally show \cite{footnoteKK} that at
intermediate to high energies, $\vert \partial \Sigma^\prime/\partial\omega
\vert \ll 1$, 
so that $\zeta(\omega) \approx 1$. Then the corresponding
high energy $1 / v_F^{\rm high} = dk/d\omega$ 
is given by
\be
v_F^{\rm high} \approx v_F^0 + \partial\Sigma^\prime/\partial k, 
\label{vfhigh2}
\ee
which together with eq.~(\ref{vflow}) leads to
the result eq.~(\ref{vfhigh}) stated in the Introduction. 
Here $\partial\Sigma^\prime/\partial k$ is an $\omega$-{\em independent
constant} as discussed in connection with the necessary conditions 
(Kramers-Kr\"onig transform of eq.~(\ref{constraint})) for
Lorentzian lineshapes.

The above results are completely general,
and we next illustrate this formalism 
for the low and high energy  dispersion of nodal QPs 
in the cuprates.

\medskip
\ni
{\bf Low energy properties $Z(x)$ and $v_F^{\rm low}(x)$:}
In Ref.~\cite{paramekanti01} we have presented a $T=0$ variational 
theory, building on early ideas of Anderson \cite{anderson}, for
the strongly correlated d-wave SC and its
low-lying excitations, examining how they evolve as a function
of hole doping $x$ from a Fermi liquid state at overdoping 
$x \gtrsim 0.35$ to a Mott insulator at half-filling ($x=0$).
Here we summarize specific results from
Ref.~\cite{paramekanti01} which relate to doping dependence
of the nodal QP spectral weight $Z(x)$ and Fermi velocity
$v_F^{\rm low}(x)$, and describe their implications 
for $\Sigma^\prime(\bk,\omega)$.

We begin by emphasizing that these results for the nodal QP
were {\em not} obtained by working with a variational
excited state, e.g., a Gutzwiller projected Bogoliubov QP.
Given the very broad linewidth of the spectral function, it is
difficult to use simple variational excited states to obtain
useful results on the coherent QP piece; see
Ref.~\cite{paramekanti03}. Instead we focused
on the {\em singularities} as a function of $\bk$ in the 
energy-integrated moments of the spectral function to extract 
information related to gapless QPs. 
In particular, $Z$ was extracted from the jump
in $\int_{-\infty}^0 d\omega A_{\bk,\omega} = n(\bk)$ 
and $v_F^{\rm low}$ from the slope discontinuity of 
$\int_{-\infty}^0 d\omega \omega A_{\bk,\omega}$ at $\bk = \bk_F$
\cite{paramekanti03}.

We found that \cite{paramekanti01} $Z(x)$ decreases with underdoping,
with $Z \sim x$ \cite{rajdeep}
as $x \to 0$, and remarkably that $v_F^{\rm low}(x)$ is essentially doping
independent and remains finite as $x \to 0$.
Using eq.~(\ref{residue}) one must then conclude from the calculated
$Z(x)$ that $\vert\partial\Sigma^\prime/\partial\omega\vert \sim 1/x$
as $x \to 0$. Further, using eq.~(\ref{vflow}),
the calculated $v_F^{\rm low}(x)$ implies
a compensating divergence 
$\partial\Sigma^\prime/\partial k \!\sim\! Ja/x$,
where the superexchange $J$ and lattice constant $a$
enter on dimensional grounds.

In the $x \to 0$ limit one thus obtains $Z \sim x$, with a renormalized
low energy Fermi velocity
$v_F^{\rm low}(x \to 0) = C J a$, where $C$ is a dimensionless constant
of order unity. Remarkably these results are {\em independent} of
the bare band-structure dispersion and thus ``universal''. From our 
numerical results \cite{paramekanti01} we find $C \approx 4.5$.
We emphasize that these non-trivial results for the doping
variation of $Z$ and $v_F^{\rm low}$ are non-perturbative, strong
coupling results which come from a variational calculation which
properly takes into account the strong local Coulomb repulsion. 

\medskip
\ni
{\bf High Energy Dispersion $v_F^{\rm high}(x)$:}
Combining all the results described above, we see from
$v_F^{\rm high} = v_F^{\rm low}/Z(x)$,
that we expect $v_F^{\rm high} \gg v_F^{\rm low}$ and also
that it will show considerable doping dependence and
increase with underdoping, given the predicted $Z(x)$.
Moreover, using eq.~(\ref{vfhigh2}) and the form of 
$\partial\Sigma^\prime/\partial k$ derived above,
it immediately follows that 
there is $Ja/x$ divergence in $v_F^{\rm high}$, 
which at small enough $x$ is much larger than and 
independent of the ``bare'' band-structure $v_F^0$.

\begin{figure}
\begin{center}
\vskip-2mm
\hspace*{0mm}
\centerline{\fig{2.8in}{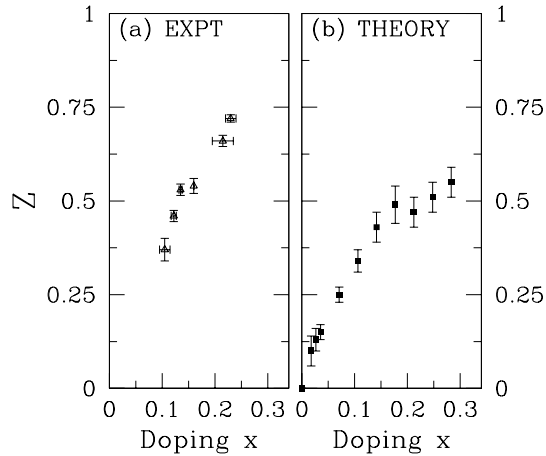}}
\vskip-6mm
\caption{{\bf (a)} Doping dependence of the nodal QP 
weight $Z$ in Bi2212 extracted from data
in Ref.\cite{johnson01} using $Z=1/(1+\lambda)$.
The doping $x$ was calculated from their sample $T_c$ using the 
empirical formula 
\cite{empiricalTc}: $86 (x-0.16)^2 = (1-T_c/T^{\rm max}_c)$, with 
$T^{\rm max}_c = 91K$ for their samples. 
{\bf (b)} $Z(x)$ predicted from a variational
approach shows that $Z\to 0$ on underdoping, and 
is in good agreement with the experimental trend and values.}
\label{fig:Z}
\vskip-6mm
\end{center}
\end{figure}

\medskip
\ni
{\bf Comparison with ARPES Experiments:}
We are now in a position to compare the theoretical results 
derived above with existing ARPES data. In Fig.~1 we plot the
nodal QP weight $Z$ as a function of hole doping $x$.
Fig.~1(a) shows $Z$ extracted from the experimental data 
of Ref.~\cite{johnson01} on Bi2212. These authors plot
$\lambda=-\partial \Sigma^\prime/\partial\omega$ as a function 
of doping, from which we obtain
$Z$ using $Z = 1/(1 + \lambda)$. In Fig.~1(b) we plot the 
nodal $Z(x)$ obtained from our variational calculation
\cite{paramekanti01} and find very good
agreement \cite{errorfootnote} between theory and experiment. 
We should note that
to leading order in $J/t$ \cite{rajdeep} the calculated
$Z(x)$ is independent of the choice of input parameters
in our theory \cite{footnote}. 
The nodal $Z(x)$ has also been reported
for LSCO \cite{yoshida02}, but in arbitrary units, which makes a
quantitative comparison difficult, however the 
trend is qualitatively similar to our theoretical prediction.

We next show in Fig.~2 the low energy Fermi velocity
$v_F^{\rm low}$ for nodal QPs
in units of $eV$-$\AA$ as a function of doping.
The experimental results shown in Fig.~2(a) are obtained from
the MDC dispersion at $E_F$ for LSCO 
(open triangles from Ref.~\cite{zhou03})
and for Bi2212 (open squares from Ref.~\cite{johnson01}).
In Fig.~2(b) we show the ``bare'' $v_F^0$ (dashed line),
which has a weak doping dependence coming from the change of the 
chemical potential in the bare tight-binding model \cite{footnote}.
The renormalized low energy Fermi velocity  $v_F^{\rm low}$
is however found to be smaller and nearly doping independent, within the 
error bars
of the theoretical calculation \cite{errorfootnote}, for the hole doping 
range $x \le 0.2$, its scale being set by $Ja$ as explained in detail above.

\begin{figure}
\begin{center}
\vskip-2mm
\hspace*{0mm}
\centerline{\fig{2.8in}{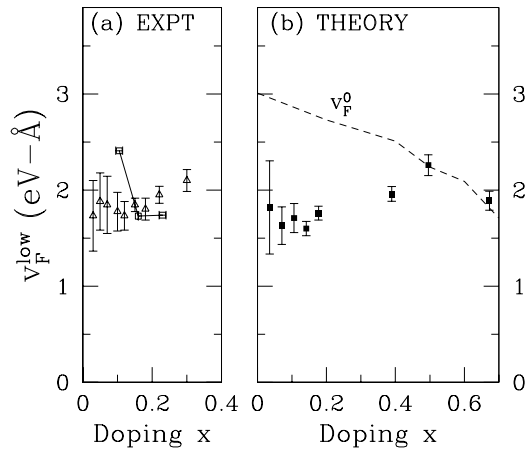}}
\vskip-6mm
\caption{{\bf (a)} The Fermi velocity 
$v_F^{\rm low}$ in Bi2212 (open squares joined by a line
as guide to the eye, Ref.\cite{johnson01})
and LSCO (open triangles, Ref.\cite{zhou03}) is 
nearly doping independent with a similar value in various 
cuprates (see Ref.\cite{zhou03} for additional data).
{\bf (b)} Theoretically predicted $v_F^{\rm low}$
using a variational approach
is constant for $x\! \lesssim\! 0.2$, independent of the bare
band structure (``universal''), and crosses over to the bare 
Fermi velocity ($v_F^0$, dashed
line) at large $x$. The values and trend are in
good agreement with experiment \cite{zhou03}.}
\label{fig:VLOW}
\end{center}
\vskip-6mm
\end{figure}

\begin{figure}
\begin{center}
\vskip-2mm
\hspace*{0mm}
\centerline{\fig{2.8in}{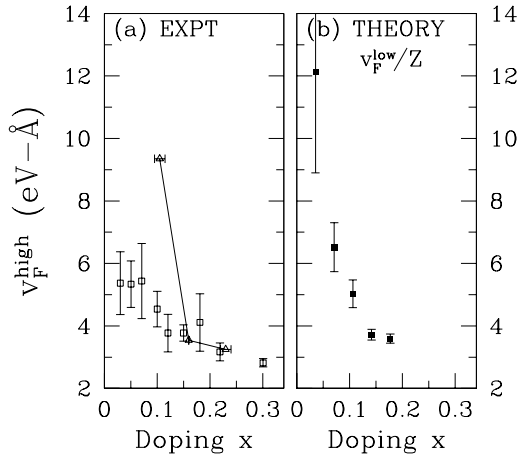}}
\vskip-6mm
\caption{{\bf (a)} Doping dependence of the high energy
energy velocity $v_F^{\rm high}$ above the kink in the MDC dispersion
from data on Bi2212 (open squares joined by a line as guide to the
eye, Ref.\cite{johnson01}) and LSCO (open triangles, Ref.\cite{zhou03}).
{\bf (b)} A variational 
calculation \cite{errorfootnote} of
$v_F^{\rm high} = v_F^{\rm low}/Z$ shows that it increases on underdoping
in agreement with the experimental trend and values for 
$x \gtrsim 0.05$.}
\label{fig:VHIGH}
\end{center}
\vskip-6mm
\end{figure}

Finally we come to the comparison of theory and experiment for the
high energy dispersion $v_F^{\rm high}$ in Fig.~3.
The data in Fig.~3(a) are obtained from
the high energy MDC dispersion for LSCO
(open squares from Ref.~\cite{zhou03})
and for Bi2212 (open triangles from Ref.~\cite{johnson01}).
In Fig.~3(b) we plot the theoretical estimate of the
high energy velocity using the expression derived above
$v_F^{\rm high} = v_F^{\rm low}/Z(x)$. We find good agreement
with the data over the range of dopings from $ x \gtrsim 0.05$
both in terms of magnitude and overall trend. There is a $1/x$
divergence in the theoretical value at small $x$, for reasons 
explained above,
but evidence for that in the very low doping data in LSCO 
is not clear.
Quite independent of our variational calculation, the general
result $v_F^{\rm high} = v_F^{\rm low}/Z(x)$ derived above, together with 
the experimentally observed behavior of $v_F^{\rm low}$ 
(constant \cite{zhou03}) and of $Z(x)$ (vanishing as $x \to 0$
\cite{yoshida02}) in LSCO, implies a  $v_F^{\rm high}$ that 
increases strongly as $x\to 0$. We hope future experiments will
clarify this situation.

\medskip
\ni
{\bf Implications for the kink:}
The success of the theory described above leads to
the obvious question: What is the origin and energy scale of the kink?
By the "kink" we mean the sharp change in the observed MDC dispersion
separating the low and high energy regimes which
occurs at an energy scale of about 70 meV \cite{lanzara01,johnson01,
zhou03}. Its origin is controversial at present, with
electrons interacting with either an optical phonon \cite{lanzara01}
or with the neutron
resonance mode \cite{norman,abanov} being the two likely scenarios.

We have presented here a framework for understanding the
low and high energy limits MDC dispersion of eq.~(\ref{dispersion}), 
in limits where the factor of $\zeta(\omega)$
goes to $Z$ and unity respectively. This general framework cannot
directly address
precisely how and at what energy scale there is a transition from one
limiting behavior to the other. The variational approach only permitted
a calculation of the low energy behavior of the 
nodal QPs --- its implications for $v_F^{\rm high}$
hinged upon the remarkable relationship between this high energy
quantity and and the low energy $v_F^{\rm low}$ and $Z$, which could
be calculated. 
However, we emphasize that the
variational calculation, which focuses on the strong Coulomb interactions,
is able to give qualitative and semi-quantitative insights into the
doping dependence of both the low and high energy dispersion of the
nodal QPs. Thus it would be very surprising if the intermediate 
energy scale kink were not also dominated by strong electron-electron 
interaction effects.
 
\medskip
\ni
{\bf Conclusions:}
To summarize, we first obtained the conditions, eq.~(\ref{constraint}) and its  
Kramers-Kr\"onig transform, for a Lorentzian MDC lineshape, and then
used this to derive the MDC dispersion eq.~(\ref{dispersion}) and
the linewidth, both of which involve the $\bk$-dependence
of the self energy in an essential way. We next showed how the
high energy dispersion was related directly via eq.~(\ref{vfhigh}) to the 
QP spectral weight $Z$ and the low-energy Fermi velocity
$v_F^{\rm low}$. We then illustrated this general formalism by using 
variational results for the nodal QP in the cuprates,
where we showed that the $Z(x)$ was strongly doping dependent and
vanishing as $x \to 0$, $v_F^{\rm low}$ was essentially doping
independent, and $v_F^{\rm high} = v_F^{\rm low}/Z(x)$.
All of these results were shown to be in good agreement with ARPES data.
 
\medskip
\ni{\bf Acknowledgments:} 
We thank J. C. Campuzano, J. Fink, A. Fujimori, P. D. Johnson,
A. Kaminski, A. J. Leggett and Z. X. Shen for stimulating discussions.
MR and NT acknowledge support through 
DOE grant DEFG02-91ER45439 and DARPA grant N0014-01-1-1062;
AP was supported by NSF grants DMR-9985255 and PHY99-07949, and
the Sloan and Packard foundations.


\begin{thebibliography}{999}
\bibitem{footnoteKK}
We have checked this over a wide range of $\omega$ using 
phenomenological forms for $\Sigma^{\prime\prime}(\bk,\omega)$ 
and Kramers-Kr\"onig analysis.
\bibitem{errorfootnote}
The error bars in the theoretical
estimates for $Z$ and $v_F^{\rm low}$ are mainly due to a combination
of limited resolution on a finite lattice in the numerics and from 
fits to the computed spectral function moments; see Ref.~\cite{paramekanti03} 
for details. 
The error in $v_F^{\rm high}=v_F^{\rm low}/Z$ is a combination
of these errors and grows with underdoping since $Z \to 0$.
\bibitem{footnote}
The input to the variational calculation is minimal: the bare nearest
and next-nearest neighbor hoppings are $t=300 meV$ and $t^\prime=t/4$ 
respectively (motivated by electronic structure calculations) and the 
superexchange $J = 100 meV$ (motivated by neutron and Raman scattering
experiments). For details, see Ref.~\cite{paramekanti03}.
\end{thebibliography}
\end{document}